\documentstyle[epsfig]{aipproc2}
\begin{document}
\title{Stochastic GW Backgrounds \\
and Ground Based Detectors }

\author{Massimo Giovannini}
\address{Institute for Theoretical Physics, Lausanne University,
CH-1015, Dorigny, Switzerland}

\maketitle

\begin{abstract}
The interplay between different ground based detectors and stochastic 
backgrounds of relic GW is described. A simultaneous 
detection of GW in the kHz and in the MHz--GHz region can 
point towards a cosmological nature of the signal. The sensitivity
of a pair of VIRGO detectors to string cosmological models is presented. 
The implications of microwave cavities for stochastic GW backgrounds
 are discussed.   
\end{abstract}

\section*{Motivations}
The purpose of the present contribution 
\footnote{Contribution to CAPP-2000, Verbier (Switzerland) July 2000. 
To appear in the Proceedings ( American Institute of Physics publication).} 
is to summarise some general results 
whose implications can improve our understanding of 
cosmological models through the next decade. In this sense, the present 
session (devoted to GW) is complementary to the one on Cosmic Microwave 
Background (CMB) anisotropies. CMB experiments are the present 
of experimental cosmology, GW represent a foreseeable future.

Every departure of the background geometry from a radiation 
dominated evolution produces GW \cite{gr}. This simple fact can be 
easily understood because, during a radiation dominated 
phase, the evolution equation of the fields describing the two polarisation 
of a GW in a (spatially flat) Friedmann-Robertson-Walker (FRW) 
background are conformally invariant. The amplitude of the detectable signal 
depends not only upon the specific theoretical model but also upon the 
specific GW detector. 

The GW spectrum ranges  over thirty decades in frequency.  
GW with (present) frequencies around
 $f_{0}\sim 10^{-18} ~h_0$ Hz correspond to a 
wave-length as large as the present Hubble radius [$h_0$  
represents the indetermination of the (present) value of the Hubble parameter]. 
For these 
waves ideal detectors would be CMB experiments.

Frequencies of the order of $10^{-4}$ Hz correspond roughly to the operating 
region of the space-borne interferometer (LISA) which will be (hopefully) 
operating at  some moment after 2017. 
Between few Hz and $10$ kHz is located the operating window of ground based 
interferometers. The (narrow) band of 
resonant mass detectors is around the kHz. 
Finally between few MHz and few GHz microwave 
cavities can be used as GW detectors. 

Between $10^{-18}$ Hz and $10$ kHz there are, 
roughly, 22 decades in frequency. 
The very same frequency gap, if applied to the well known electromagnetic 
spectrum,  would drive us from low-frequency radio waves up  to 
x-rays or $\gamma$-rays. As the physics explored by radio waves is very 
different from the physics probed by $\gamma$ rays 
it can be argued that the informations carried by low and high 
frequency GW must 
derive from two different physical regimes of the theory.
 
In particular, low frequency GW are sensitive to the large scale features 
of the given cosmological model and of the underlying theory of gravity, 
whereas high frequency GW are sensitive to the small scale features of a 
given cosmological model and of the underlying theory of gravity. 
For instance string theory is expected to lead to a description of gravity 
which resembles very much Einstein-Hilbert gravity at large scales 
but which can deviate from Einstein-Hilbert gravity at smaller scales. 
That is only one of the many reasons why it is very important to have 
GW detectors operating over different frequency bands.  

\section*{Ground based GW detectors}
GW detectors can be divided in three broad classes: resonant 
mass detectors, interferometers and microwave cavities.
There are five (cryogenic) resonant mass detectors which are 
now operating: NIOBE \cite{niobe} (Perth, Australia), 
ALLEGRO \cite{allegro} (Baton Rouge, Lousiana, USA), AURIGA 
\cite{auriga} (Legnaro, Italy), EXPLORER \cite{explorer}
(Geneva, Switzerland) and NAUTILUS \cite{nautilus} 
(Frascati, Italy). They all have 
cylindrical shape (the are ``bars''). They are all made in Aluminium 
(except NIOBE which is made of Niobium). Their approximate mass is of the 
order of $2200$ kg (except NIOBE whose mass is of the order of $1500$ kg).
Their mode frequencies range from $694$ Hz (in the case of NIOBE) to 
the $912$ Hz of AURIGA.
The shape of a resonant mass detector does not need to be cylindrical. 
In particular the nice idea of spherical GW detectors 
is being actively pursued \cite{sph,sph2}.  

There are, at the moment, four Michelson-Morley interferometers 
being built. They are 
GEO \cite{geo} (Hannover, Germany), TAMA \cite{tama} (Tokyo, Japan), 
VIRGO \cite{virgo,virgo2} (Cascina, Italy),
and the two LIGO \cite{ligo} 
(in Hanford [Washington], and Livingston [Lousiana], USA).
The arms of the instruments range from the 400 m of TAMA up to the 
three km of VIRGO and to the 4 km of LIGO. The effective optical path of 
the photons in the interferometers is greatly enhanced by the use of 
Fabry-P\'erot cavities.

Microwave cavities have been originally proposed as GW detectors in the 
GHz--MHz region of the spectrum \cite{mw1}. A first prototype has been 
built in MIT in 1978 showing that this idea could be 
actually implemented in order to detct small harmonic displacements \cite{mw2}. 
It is not unreasonable to think that sensitive measurements could 
be performed in the near future. In particular improvements in the quality 
factors of the cavities (if compared 
with the prototypes of \cite{mw1}) could be foreseen.
Two experiments (in Italy \cite{paco} and in England \cite{bir})
 are now trying to achieve this goal with slightly different technologies.
  
\section*{Stochastic GW backgrounds}

Define 
\begin{equation}
\Omega_{{\rm GW}}(f,\eta_0)\,=\,\frac{1}{\rho_{c}}\,
\frac{{\rm d} \rho_{{\rm GW}}}{{\rm d} \ln{f}}
\label{Omegath}
\end{equation}
as the fraction of critical energy density stored in relic GW at 
the present time.
In ordinary inflationary models $\Omega_{\rm GW}(f)$ is minute.
Consider an ordinary inflationary phase being replaced 
by a radiation dominated phase which evolves, in its turn,  
into a matter dominated epoch of expansion. Then, the  GW spectrum 
has two branches: a soft branch (between $10^{-18}$ and $10^{-16}$ Hz) and a 
quasi-flat branch for $f > 10^{-16}$. In the soft branch the logarithmic 
energy spectrum decreases as $ f^{-2}$. The gross features of inflationary 
models forbid $h_0^2 \Omega_{\rm GW}$ being larger that $10^{-15}$ in the 
flat branch of the spectrum. The reason  is that 
at low frequencies (of the order of $f_0$) the COBE bound imposes 
$h_0^2\Omega_{\rm GW} < 7\times 10^{-9}$. If we now assume (rather 
optimistically, indeed) that the inflationary GW spectrum 
is flat for $f > 10^{-16} $ Hz we can compute quite easily the signal-to-noise
ratio (SNR) for a pair, say, of correlated interferometers. Thus, 
even with a pair of advanced devices we can get (at most) down to 
$\Omega_{\rm GW}(0.1 ~{\rm kHz}) \sim 6.5 \times 10^{-11}$ \cite{ar,noi,pol}. 
Thus, ordinary 
inflationary models are invisible by ground based detectors. 

In order to have a large detectable signal for frequencies larger than 
$10^{-16}$ Hz we have to invoke departures from scale invariance \cite{m1,m2}, 
i.e. 
scaling violation which can take place both in the case of quintessential 
inflationary models \cite{pv,m3} and in the case of string cosmological models
\cite{ven,m4}.    
The scaling violations should go in the direction of logarithmic energy 
spectra which increase in frequency. Only in this case 
a large signal can be expected at high frequencies \cite{m5}. 

By recalling that interferometers will be operating between few 
Hz and $10$ kHz and by recalling that microwave cavities will instead be 
operating for frequencies higher than the MHz we can envisage two different
theoretical situations \cite{noi}. We can think of 
a model where the signal at the frequency of the interefreometers 
is small. However, thanks to the frequency growth the same model 
could lead to a large signal at the frequency of the 
microwave cavities. We could also have the situation where 
the signal is large both at the interferometers 
scale and at the scale of the microwave cavities. The first 
case corresponds to quintessential inflationary models \cite{m2,m3} 
the second case 
corresponds to string cosmological models \cite{m4,m5}. 

In both cases a detection of a signal at the scale of the microwave cavities 
would be a probe of the cosmological nature of the process producing the GW
 background: no astrophysical sources are expected at such high frequencies. 
In both cases there are frequency regions where the signal exceeds ( even by 8 
decades in $\Omega_{\rm GW}(f)$ ) the inflationary prediction.
The idea of using electromagnetic detectors in order to probe 
stochastic GW backgrounds at high frequencies 
was also pointed out Grishchuk \cite{gr2}.

\section*{Sensitivity of a VIRGO pair to string cosmological gravitons}

Recently there have been concrete steps \cite{gia} towards the proposal of building 
in Europe an interferometer of dimensions comparable with VIRGO \cite{virgo}. In view of 
this idea the sensitivity of the correlation between two VIRGO-like detectors 
to a generic stochastic GW background has been scrutinised in a number of papers
\cite{noi,noi2}. 

Different locations for the site of the second detector 
have been examined in light of the possible stochastic sources.
Given a theoretical model whose signal is, in principle, large 
enough to be detected by such a device there are two important aspects. In the first place, 
assuming a specific configuration of the VIRGO pair, we ought to know how the detectable signal 
 changes by changing the parameters of the model. Secondly, we would like to know how 
much of the parameter space of the model can be probed assuming that the 
noises of the detctors are reduced by a given amount. These two questions have been 
discussed in \cite{noi2}.
\begin{figure}[!hb]
\vspace*{-3.5cm}
\centerline{\epsfxsize = 21.5 cm  \epsffile{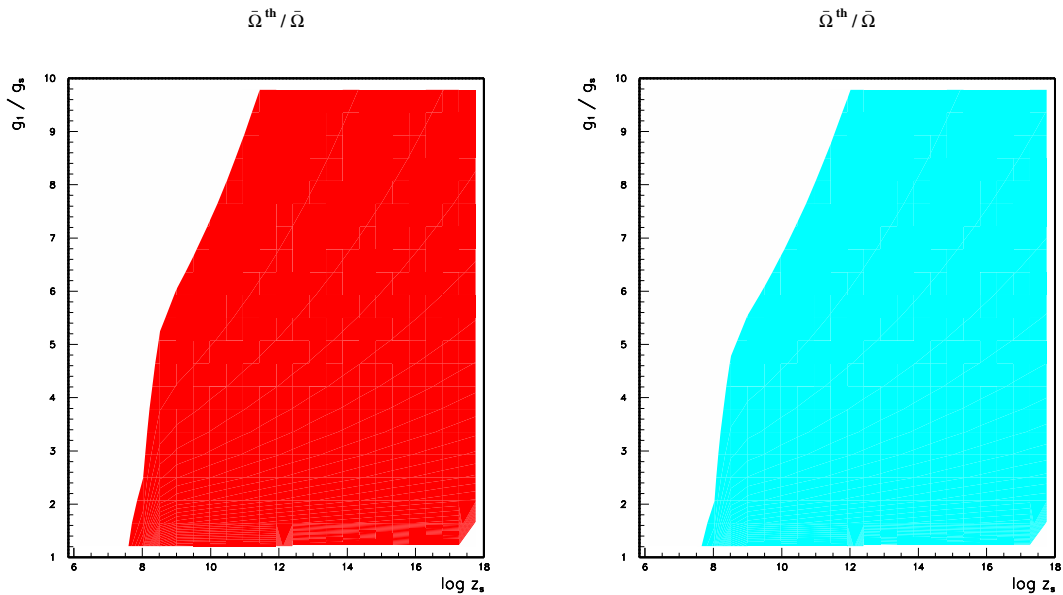}}
\vspace*{-18.0cm}
\caption[a]{Under the assumption of selective thermal noise reduction
we illustrate the behaviour of visibility region of the VIRGO pair in the 
case on growing logarithmic energy spectra of string cosmological type.
 At the left the overlap reduction is taken in the case where the two detectors 
are separated by $56$ km. At the right the distance between the two 
sites is taken to be $958.2$ km.}
\label{strrid}
\end{figure}
In Fig. \ref{strrid} the visibility region of a VIRGO pair is illustrated 
in terms of the parameter space of string cosmological models. The 
shaded areas correspond to regions of the (two-dimensional) 
parameter space giving a signal-to-noise ratio larger than (or equal to) one. 
In both plots the normalisation the signal is compatible with 
all the physical bounds we can apply to stochastic GW backgrounds \cite{m3}.
Fig. \ref{strrid} refers to the case where the noises 
induced by the pendulum and by its internal modes are suppressed, respectively, 
by a factor of $10$ and by a factor of $100$ if compared with 
the initial VIRGO noise power spectrum. Different noise 
reductions give rise to different visibility regions.
$\overline{\Omega}$ is the minimal detectable $\Omega_{\rm GW}(f)$ 
at a frequency of $0.1$ kHz after one year of correlation of the two detectors.
$\overline{\Omega}^{\rm th}$ is the theoretical normalisation 
of the spectrum. For all the points in the shaded regions 
$\overline{\Omega}^{\rm th}/\overline{\Omega}>1$.

\section*{Conclusions}

It is very important to have detectors 
in different frequency regions. Interferometers and resonant 
mass detectors on one hand and microwave cavities on the other hand 
are complementary devices.  Arrays of detectors at intermediate and 
high frequencies (i.e. larger than the MHz) can provide 
 informations on the spectrum. Large signals coming from 
string cosmology and quintessential inflation can be, in this context, 
inspiring for the experimental work.

\end{document}